# Structural Phase Transitions

R. A. Cowley
*Department of Physics, Oxford University*
*Clarendon Laboratory, Parks, Road*
*Oxford, OX1 3PU, UK*

S. M. Shapiro
*Condensed Matter Physics/Materials Science Department*
*Brookhaven National Laboratory*
*Upton, NY 19973, USA*

**Abstract**

Gen Shirane began studying ferroelectrics while he was still based in Japan in the early 1950's. It was therefore natural that when he arrived at Brookhaven and began specialising in neutron scattering that he would devote much of his energy and expertise studying structural phase transitions. We review the ground breaking experiments that showed the behaviour of antiferroelectrics and ferroelectrics were reasonably described in terms of the soft mode concept of structural phase transitions. This work lead directly to Gen being awarded the Buckley prize and, jointly with John Axe, awarded the Warren prize. We then describe his work on incommensurate phase transitions and in particular how the giant Kohn anomalies are responsible for the structural instabilities in one-dimensional metals. Finally Gen carefully investigated the central peak and the two-length scale phenomena that occur at most if not all transitions. Due to Gen's elegant experimental work we know a great deal about both of these effects but in neither case is theory able to explain all of his results.

**1. Introduction**

It has been a pleasure for us to write this article as a tribute to the enormous impact Gen Shirane had on our scientific and personal lives. In this article we shall try to describe the important work that he performed on structural phase transitions. Before he came to Brookhaven in 1965 he had worked largely on materials that were ferroelectric. His work[1] started in Japan and he made a number of x-ray and neutron diffraction experiments[2-5] particularly while he was in the crystallographic laboratory of Pepinsky. He also wrote together with Jona his extremely clear and logical book on ferroelectricity[6]. This topic was his primary interest before he joined the staff at Brookhaven and learnt about neutron scattering in detail. His neutron scattering career began with studies of magnetism and some phonon experiments but it was natural for him to apply neutron scattering to structural phase transitions as soon as he was able to conduct his own experiments. In 1973, he was awarded the Oliver E. Buckley Solid State Physics Prize and co-recipient with John Axe of the Warren Award for his innovative experiments on structural phase transitions[7] showing that this subject was both his first scientific love and was the first area in which he made a large impact. At the end of his life in 2005, he was still actively studying structural phase transitions in diffuse ferroelectrics such as $PbMn_xNb_{1-x}O_3$, PMN, showing that the topic was not only his first scientific love but also his last love.

In this article we shall attempt to review the impact that Gen's experiments and intuition have had on the field of structural phase transitions. Throughout the review we shall concentrate on the properties of strontium titanate. This is because strontium titanate is one of the most studied structural phase transitions and it illustrates very clearly what we now know about structural phase transitions and what is not yet properly understood. Gen did experiments on this material spasmodically over nearly 40 years and it is very instructive to examine how his and our knowledge steadily progressed to the present day. The next section is broken into two parts. The experiments described are ones that were motivated by the Landau theory of phase transitions and the soft mode concept and the first part describes the largely successful application of these ideas to antiferroelectric transitions in materials like strontium titanate. The second half of this section describes the application of the same ideas to ferrodistortive transitions. Here the theory works less well but the results are in general accord with the soft mode theory. This section is largely a review of the experiments performed by Gen on structural phase transitions and for which he was awarded the Oliver Buckley Prize and the topics are also summarised in his Buckley lecture[7]. In section 3 we describe incommensurate phase transitions and phase transitions in one-dimensional metals. Both of these topics are less well understood but the extent to which they too can be described by soft mode theory are described as well as having aspects that are more difficult to understand.

In section 4 we describe the experiments that showed that there were two-time scales associated with many phase transitions. This is in striking contradiction with the soft mode theory and with more recent calculations using renormalisation group theory. The experiments are a direct result of using excellent neutron scattering instrumentation and developments in new instrumentation and have now been extended to magnetic phase transitions between two fully ordered phases. In our view the results showing two-time scales are not yet understood, although the experimental results were performed over 30 years ago. Section 5 is concerned with the two-length scale phenomena observed now not only at structural phase transitions but also at magnetic phase transitions when scattering experiments are performed with high resolution and in particular with x-ray scattering techniques. It has been shown that this behaviour is associated with the surface but there is not yet a satisfactory theory for this effect.

Gen was an experimentalist. After coming to Brookhaven he learnt about neutron scattering and became an authority in the use of the triple axis crystal spectrometer shown in fig. 1. This instrument was developed by Bert Brockhouse in the mid 1950's at the Atomic Energy of Canada. It consisted of a monochromator which produced an incident beam of nearly monochromatic neutrons of mean energy $E_0$ and wave-vector $\mathbf{k_0}$, a sample table which could both carry the sample environment and enable the crystal orientation to be controlled and a single crystal analyser that selected neutrons with mean energy $E_1$, and wave-vector $\mathbf{k_1}$. Conservation of energy and wave-vector then enables the frequency ω, and the reduced wave-vector of the excitations, q, to be determined:

$$E_0 - E_1 = \hbar\omega$$
$$k_0 - k_1 = Q = \tau \pm q \quad (1)$$

where τ is a reciprocal lattice vector.

In the 1950's this instrument was difficult to use and the experiments took a long time and there was strong competition from time-of flight neutron scattering techniques. A major step forward was the realisation that the instrument could be controlled so as to perform scans in which the wave-vector transfer was held constant while the energy transfer was varied and vice versa. This required the use of a computer to calculate the spectrometer angles and became completely effective only when computers became cheap enough that each instrument could have a dedicated computer. This development occurred around the time that Gen joined Brookhaven and led to a dramatic increase in the effectiveness of the spectrometers.

The monochromators used to produce the incident beam of neutrons and used as analysers were initially made from single crystals of aluminium, copper or lead. The crystals were grown in various ways to try to optimise the scattered neutron flux but only Bragg scattered about 20% of the theoretical number of neutrons. They also had the disadvantage that the Bragg scattering was governed by

$$n\lambda = 2d\sin(\theta_B) \quad (2)$$

so that a given set of planes reflects neutrons with wavelengths λ, λ/2, and λ/3. This was very inconvenient because the energy transfer of a peak was not then well defined because of this order contamination. The problem was solved essentially by Gen and his collaborators who used pyrolytic graphite firstly as an efficient mosaic monochromator with vertical focussing and secondly as a particularly efficient filter for 15 meV neutrons, because the graphite then Bragg reflected the second and third orders out of the neutron beam. Although not all experiments use pyrolytic graphite, its introduction has made an enormous difference to the field with the efficiency of the monochromators and the elimination of the order contamination made the interpretation of the results much easier. The effectiveness of the triple axis spectrometer was probably increased by about a factor of five - ten directly through the introduction of pyrolytic graphite and the development of powerful personal computers. Gen was an expert exponent in the use of the triple axis spectrometer and he and his collaborators developed the detailed techniques so that triple-axis spectrometers dominated inelastic neutron scattering from single crystals for the next 30 years. The techniques developed are described in the book 'Neutron Scattering with a Triple Axis Spectometer'[8] which was written in collaboration with John Tranquada and one of us.

## 2. The Classical Approach to Structural Phase Transitions

### 2.1 Antiferrodistortive Transitions

The classical theory of phase transitions is built upon the Landau theory of expanding the free energy in a power series[9] developed in 1937. A simplified form of this theory can be obtained if the displacement occurring at a phase transition has three components of a vector Q namely $Q_x$, $Q_y$, and $Q_z$, then the free energy can be written in terms of the vector as far as the quartic terms for a cubic system as;

$$F - F_0 = aQ^2 + uQ^4 + v(Q_x^4 + Q_y^4 + Q_z^4) + ... \qquad (3)$$

where *a* is assumed to be temperature dependent $a=A(T-T_c)$ and *v* determines the direction of the ordered moment which is along the [100] directions if *v* is negative and along the [111] direction if *v* is positive.

Structural phase transitions are associated with a dynamical instability by the soft mode theory put forward by Cochran[10] although there had been earlier similar proposals by Raman and by Anderson. The soft mode concept is that on cooling a material from a temperature above the transition temperature, a normal mode of vibration of the crystal decreases to zero frequency when the crystal becomes unstable and distorts to a new structure. Cochran suggested on the basis of Landau theory and the anharmonic interactions between the phonons that the temperature dependence of the soft mode would be:

$$\omega_S^2 = K(T - T_c) \qquad (4)$$

After this approach was put forward a number of experiments were performed to show the correctness of the ideas but these did not give a conclusive proof of the concept until experiments were performed on the zone boundary mode of $SrTiO_3$. Initially the distorted structure of this material was proposed by Unoki and Sakudo[11] on the basis of their ESR measurements as illustrated in fig.2. The high temperature structure is the undistorted perovskite structure with a cubic unit cell and the Sr atoms at (0,0,0), the Ti at (1/2,1/2,1/2) and the O atoms at (1/2,1/2,0), (1/2,0,1/2) and (0,1/2,1/2,1/2). As shown in fig. 2, the distorted low temperature structure has the oxygen octahedra rotated about one of the cube axes below the transition temperature of about 110K. The soft mode was measured by Raman scattering techniques by Fleury et al[12] and then detailed and independent neutron scattering measurements were made both at Chalk River[13] and at Brookhaven[14-15]. The soft mode in this case is triply degenerate and occurs at the R point (1/2,1/2,1/2) in the Brillouin zone, while the eigenvector is determined totally by symmetry. The frequency of this mode was determined above $T_c$ and typical results are shown in fig 3 while the temperature dependence of the results is shown in fig. 4. The results are clearly in very reasonable agreement with the soft mode theory as given above. Below the transition the quartic terms in the free energy determine the ordering direction as described for the Landau theory. In $SrTiO_3$ the Landau parameter $v$ is negative and the distortion is along the [100] axes. The triply degenerate soft mode is split by the distortion into a singlet mode with the rotation axis parallel to the static distortion and a doublet mode with the rotation axes perpendicular to the distortion, as shown in fig.5. Figure 6 shows a part of the phonon dispersion curves of $SrTiO_3$ and the soft mode only softens appreciably in the vicinity of the (1,1,1) zone boundary. The eigenvector of the soft mode was measured by measuring the scattered intensity of neutrons at several equivalent but different zone boundaries. The results showed conclusively that the mode was that identified by Unoki and Sakudo.

Other phase transitions are similar to that in $SrTiO_3$. The transition in $LaAlO_3$ occurs at a much higher temperature of 808K and the experiments[16-17] showed that the transition is also continuous and that the soft mode picture is essentially correct. However, $v$ is positive leading to the ordered structure having a [111] distortion in which the modes rotate around all three cube axes by the same amount.

Another example of this type of phase transition occurs in $KMnF_3$ which has a structural phase transition at 186K and another one at 86K[18-20]. In this material the higher temperature transition is identical with that occurring in $SrTiO_3$ as shown in fig. 7 while the lower temperature transition is associated with a soft mode at the M point, (1/2,1/2,0), in the Brillouin zone. This latter mode is associated with rotations of the fluorine octahedra but for the M point successive planes in the distorted structure rotate in the same direction unlike when the mode is at the R point. The dispersion curve between the R and M point associated with rotations of the octahedra, fig. 2, shows very little dispersion as shown in fig. 8 and so it is not surprising that both modes become unstable at very similar temperatures.

Another example of antiferroelectric transitions is the study of tetragonal $Tb_2(MoO_4)_3$ for which the soft mode is found to be a doubly degenerate mode at the M (1/2,1/2,0) point in the Brillouin zone. The square of the soft mode frequency is linear in the difference in the temperature from $T_C$ = 434K as predicted by the soft mode theory[21-22]. The new feature comes in the low temperature phase because at the transition the structure changes from tetragonal to orthorhombic and the square of the order parameter couples with the piezoelectric strain so that the structure becomes ferroelectric with the magnitude of the ferroelectric moment dependent on the exact pattern of the displacements occurring at the M point in the low temperature phase.

**2.2 Ferroelectric Transitions**

The behaviour of the soft modes in ferroelectric materials was more complex than the behaviour of the antiferroelectric materials. Experiments[23-24] on the materials which barely become ferroelectric at very low temperatures such as $SrTiO_3$ and $KTaO_3$ did show that the q=0 optic modes were soft and temperature dependent decreasing to a low frequency at the lowest temperature. In fig. 9 we show the results from Gen's measurements[23] on $KTaO_3$. The temperature dependence of the square of the frequency of the soft mode is approximately linear in temperature and inversely proportional to the dielectric constant as predicted by the soft mode theory. More recently there have been very detailed studies of this behaviour especially when $KTaO_3$ is doped with small amounts of Li or Nb that have given very interesting results[25] but we shall not describe these investigations here.

The perovskite ferroelectrics which are cubic at high temperatures and which then have a ferroelectric transition to a distorted phase were also studied by Gen. A surprising result at the time was the observation of Comés with x-ray scattering techniques that there were planes of diffuse scattering in the cubic phase and that the planes disappeared as the material underwent its successive transitions to a tetragonal phase, an orthorhombic phase and then a hexagonal phase. Furthermore when neutron scattering measurements were made in the cubic phase of $BaTiO_3$[26-27], $PbTiO_3$[28] and $KNbO_3$[29] it was found that the lowest frequency q=0 transverse optic mode was overdamped or at least did not give rise to a well-defined neutron group in the inelastic scattering, as shown for $PbTiO_3$ in fig. 10.

As a result of the experiments by Gen and his collaborators it is now known that the anharmonic interaction between the phonons is very important in these materials and that this can distort the scattered neutron distributions so that the frequencies and intensities of the modes differ strongly in different Brillouin zones[30]. An example, shown in fig. 11, shows experimental results for $BaTiO_3$ with a coupling at finite wave vector between the lowest frequency optic and acoustic mode in the cubic high temperature phase that produces the way the intensity changes the line shapes in different domains. Another outcome of their investigations was that the phonon modes were very anisotropic and had low frequencies and high intensities along the directions where Comés et al had observed the strong x-ray scattering streaks. More problematical is the behaviour of the transverse optic mode at q=0. In all of these materials the damping of this mode increases as the wave-vector becomes smaller and it has not been possible to resolve an underdamped phonon mode in these ferroelectric materials as illustrated for example in fig. 12. The reason for this is still not understood because the same effect also occurs in the relaxor ferroelectrics that have the same crystal structure but have disorder on one of the lattice sites. Gen's experiments on the relaxors were performed 30 years after his work on the ferroelectrics and his results on these materials will be described in another article.

One of the Gen's achievements on ferroelectrics was the measurements of the eigenvectors of the ferroelectric mode at reasonably small wave vector. The intensity of the scattering depends on the displacement of the modes and on the scattering of each individual atom[31]. By measuring the scattered intensities in several Brillouin zones, Gen was able to deduce the amplitudes of the displacements in the soft mode. The displacements occurring at the phase transition were then clarified because the displacements of the soft mode did not then involve the displacement of the centre of mass of the unit cell and furthermore he could distinguish between the soft modes proposed by Last and by Slater. These measurements were made for many of the ferroelectrics discussed above.

Another example of a ferroelectric phase transition is $KD_2PO_4$. In this case the ferroelectric fluctuations are closely associated with the motion of the hydrogen bonds, the fluctuations are overdamped and all the critical scattering is quasi-elastic while the phonon modes are not appreciably temperature dependent[32]. The ferroelectric fluctuations were correctly identified by Gen and the distribution of the neutron scattering was measured in the (H0L) plane as shown by the contour map in fig. 13. At the time the butterfly shape was unexpected but Gen correctly realised that it arises because the material is uniaxial with the spontaneous polarisation below the transition occuring along the unique z-axis. The singularity arises because the Fourier transform of the dipolar forces between the ferroelectric fluctuations has a singular form $Cq_z^2/q^2$ showing that the fluctuations are small when the dipolar forces are a maximum $q_z = q$, but large when the dipolar term is absent and the wave-vector is along the x or y axes. In the experiment Gen also measured the distribution of scattered intensity in the (HK0) scattering plane. The theory of the scattering is then complicated by the linear coupling between the transverse acoustic mode governed by the elastic constant $c_{66}$ and the ferroelectric fluctuations. The resultant scattering is shown in fig. 14. The theory of this scattering in this plane was calculated by one of us and gives very reasonable agreement with this complex scattering pattern observed. This line of research was continued by measurements on the related material $CsD_2PO_4$ which has a different structure and hydrogen bond network to that in $KD_2PO_4$. This material has a one –dimensional network of hydrogen bonds along the ferroelectric b-axis[33-35]. The diffuse scattering in the (H,K,0) plane is then similar as shown in fig. 13 but much narrower in the direction parallel to the ferroelectric axis. The narrowing could be caused by much larger dipolar forces but it is more likely that it arises from the one-dimensional arrays of coupled hydrogen bonds and that the bonding rules of the hydrogen bonds, ice rules, give rise to the singularity in the scattering. This was further shown by the study at Brookhaven by Axe and Youngblood[36] on copper formate tetrahydrate that has a two-dimensional network of hydrogen bonds and shows diffuse scattering that is non-analytical in reciprocal space and that these anomalies can be explained by the 6 vertex hydrogen bond model. These models are now also used to describe the magnetic scattering in pyrochlores.

Although not a ferroelectric transition the structural phase transition in $Nb_3Sn$ is a ferrodistortive transition. In this case the soft mode is an acoustic mode that decreases in velocity as the phase transition is approached and the crystal then distorts so that the forbidden (003) Bragg reflection becomes active[37-40]. In fig. 15 we show the temperature dependence of the reflection and it increases with decreasing temperature as predicted by the Landau theory. Measurements of the slope of the acoustic phonon mode are shown in fig.16 and show a marked decrease in frequency as the transition is approached. However the measured slope does not decrease to zero in the measurements, firstly because it is impossible to make measurements at very low q vectors and secondly because the intensity in the quasi-elastic peak increases more rapidly than expected as shown in fig. 17. More examples of this type of behaviour will be described in section 4.

**3. Incommensurate Phase Transitions**

So far we have discussed systems in which the soft modes have wave-vectors at the centre or zone boundaries in the Brillouin zone. New features arise when we consider transitions that are associated with incommensurate wave-vectors in the Brillouin zone. The first of these is that the symmetry of the soft mode must be even because there will always be wave-vectors with both positive and negative signs. The second new feature is that in a strictly incommensurate transition the phase of the soft mode is undefined and the phase has a broken continuous symmetry. It is then an example of a Goldstone mode and there is always a branch of the phonon dispersion relations that corresponds to movements of the phase and which goes linearly to zero as the wave-vector becomes close to that of the incommensurate modulation. Much of the effort associated with incommensurate materials has been to identify and study the phason modes and since they are often overdamped this has proved a very difficult quest.

One of the simplest incommensurate materials, $K_2SeO_4$, was studied[41-42] by Gen. It is an insulator with a relatively low transition temperature of 128K and exhibits a lock-in transition at 90K. Figure 18 shows the low energy phonon dispersion curves propagating along the [100] direction. There is clearly a soft mode and it softens over a considerable fraction of the Brillouin zone. The square of the minimum frequency decreases as the transition temperature is approached but there does not seem to have been an accurate determination of the soft mode frequencies as a function of temperature.

The wave-vector of the distortion in $K_2SeO_4$ was measured below the incommensurate phase transition and it changed from $(1-\delta)/3$ with $\delta = 0.02\text{Å}^{-1}$ to $\delta = 0.005\text{Å}^{-1}$ at the lock-in transition of 93K. The lock-in transition is a first order transition where the $\delta$ jumps to zero. Just above the first order transition scattering was observed at several higher order positions such as $(1-2\delta)/3$. It was easier to study the inelastic scattering below the lower temperature phase transition where there was an interaction between the transverse acoustic mode and the phason mode as illustrated in fig. 19 for the intensities and frequencies. However, one of the main contributions of this research was the first detailed use of Landau theory and soft mode theory for incommensurate phase transitions including the effects of phasons and lock-in transitions.

A more complex incommensurate phase transition occurs in the one-dimensional conductor KCP, $(K_2Pt(CN)_4Br_{0.3}.3.2D_2O)$[43-46]. This and similar materials have one-dimensional chains oriented along the c-axis. These chains are conducting through the Pt ions and since there are only 0.3 Br atoms per formula unit simple band theory predicts that the one-dimensional electron band will be only 0.85 full. If there is no interaction between the chains the materials are examples of one-dimensional conductors and there will be no long range order formed above zero temperature. These materials can be expected to have large Kohn anomalies with a wave vector of $2k_F = 0.3$ of the way to the Brillouin zone along the c-axis. The phonon dispersion curves were measured, as shown in fig. 20, and the phonons are reasonably normal apart from a small region shown along the $[1//2.1/2.\zeta]$ and $[0,0,\zeta]$ directions which have very large anomalies when $\zeta = 2k_F$. The detailed scattering in the neighbourhood of these anomalies was measured a as functions of temperature and the results are shown in fig. 21 both above and below the phase transition at about 120K. Unfortunately the sharp decrease in the frequency of the dispersion curves prevents a detailed study of the dynamics in these systems while furthermore the low temperature structure is not a long range distorted structure but has at best only short range order. These results and the fact that the conduction is produced by Br ions that are arranged as defects makes it difficult to reconcile with a detailed three-dimensional theory for these nearly one-dimensional conductors. Nevertheless the measurements do show that the transition is associated with a strong Kohn anomaly at a wave-vector given by the number of defect Br ions in agreement with the simple theory.

Similar results were also obtained for the organic metal TTF-TCNQ. In this case the experiments were very difficult due to the small size of the samples, and because it was necessary to deuterate the crystals if a large background was to be avoided. The subject was also of importance to Gen because he initially competed with Herb Mook of Oak Ridge to obtain the best data. It was one of the first examples of a competition between Brookhaven and Oak Ridge which tended to become more intense and involved more experiments on topically important materials as time proceeded. In the case of the experiments[47-55] on TTF-TCNQ the experiments showed that there was a large Kohn anomaly as found for KCP although not quite so sharp in wave-vector as that in KCP and much of the effort was then directed to understanding the 3 low temperature ordered phases below 50K. The structures of these phases depend on the detail of the inter-chain interactions and the theory is sufficiently detailed that we shall not discuss it further here.

Another example of a one-dimensional material studied by Gen was the mercury chain compound $Hg_{3-\delta}AsF_6$. The structure consists of a tetragonal array of $AsF_6$ octahedra with interpenetrating chains of Hg atoms running along the [100] and [010] directions. These two aspects of the structure are only weakly coupled at least in part because the number of Hg atoms is incommensurate with the $AsF_6$ octahedra. At high temperatures the Hg atoms give rise to sheets of scattering perpendicular to the [100] and [010] directions showing that the Hg chains are independent of one another, of the $AsF_6$ octahedra and that there is no long range order[56-59]. Figure 22 shows the dispersion curve obtained at room temperature for scattering from the Hg atoms showing that in agreement with theory this scattering is linear with wave-vector difference away from the centre of the sheets. On cooling below 120K the system orders due to the three-dimensional interactions between the chains of atoms. In fig. 23 we show the intensity at two wave-vectors on the sheets: one where the Bragg reflection occurs and one where it does not. The interaction orders the different chains along the [100] direction and simultaneously the phases between the two sets of chains were also determined. This gives rise to a set of three-dimensional Bragg peaks and a fully ordered crystallographic structure. As with KCP and TTF-TCNQ these quasi one-dimensional systems have been shown either to develop full long range order above T = 0K due to the inter-chain interactions or at least short range order that gets frozen before developing full three dimensional order.

**4. Non classical behavior: central peaks**

The discovery of the "central peak" in structural phase transitions was made in Kjeller, Norway in 1971 by Tormod Riste[60] a long time collaborator of Gen. Riste had invited Gen to visit Kjeller for a summer in 1970 with the idea to look for some quasieleastic scattering in $SrTiO_3$. Riste had worked on magnetic systems for several years and noted that in magnetic systems there are soft modes (magnons below Tc) and a quasielastic peak associated with spin diffusion. Gen, however, had a different idea of what to study and dismissed Riste's notion with a wave of his hand and pursued his own idea of probing the soft mode dynamics of $SrTiO_3$ up to higher temperatures to see if it deviated from predicted behavior [15]. After Gen departed from Norway, Riste persevered with his original idea and first observed the central peak in SrTiO3 and showed the simultaneous existence of two modes in the critical region of the phase transition: the soft phonon mode and the central mode[60]. No energy linewidth of the central peak was observed and the central mode dominated the scattering close to the transition temperature. It is interesting to examine why Riste observed the central peak in $SrTiO_3$ when it was missed in both the Brookhaven[14] and Chalk River[13] experiments illustrated in fig. 3 where the scattering clearly has a large quasielastic component. Both of these experiments were conducted with triple axis spectrometers before the use of graphite filters was common. They, therefore, suffered from order contamination of $\lambda/2$ so that any elastic scattering at the R point would be greatly contaminated and so was neglected. Of course, a study of both experiments with hindsight shows that they did observe the central peak but the participants had not noticed it. Riste clearly observed it because he used a time-of-flight spectrometer which did not suffer from the order contamination problem. This disadvantage of using a triple axis instrument was overcome by the use of pyrolytic graphite filters.

Gen's interest in $SrTiO_3$ continued when he returned to Brookhaven. He was intrigued by the results of Müller and Berlinger [61] who measured the critical exponent of the order parameter of the structural phase transition and found $\beta=0.33$. This was a surprise, since these structural phase transitions were thought to be mean field like and would have $\beta=0.5$. Gen's speculation was that the transition was first order and the crystals measured were slightly strained with a smeared out transition that appeared continuous with a smaller exponent. A nearly perfect, strain free crystal was grown for the neutron experiments performed at Brookhaven's High Flux Beam Reactor. The crystal was brownish in color and had a transition temperature $T_c$=99.5 K, which is 5K less than had been reported earlier in the literature [13,14,61]. To the surprise of Gen and his collaborators, the transition in this nearly perfect crystal proved to be continuous, with β~.33, consistent with the results of Müller[61] and Riste[60]. In addition the central mode was also observed and dominated the spectra near the transition temperature. Gen continued his study of $SrTiO_3$ for 30 years with surprises continually appearing.

Figure 24 shows the high resolution spectra obtained from the special crystal prepared for the Brookhaven experiments[62]. The spectra were measured at the 1/2 (1,1,1) R-point zone boundary, which is the q-vector of the soft mode as shown in Fig. 3. The left side shows mode softening of the phonon mode and the presence of the central peak. The right side is an expansion of the energy scale and more clearly shows that the central peak diverging as the transition temperature 99.5 K approaches. This central peak was subsequently observed in many other structural phase transitions. Fig. 25 shows that it is also present in the structural phase transition in $KMnF_3$, where it is compared with $SrTiO_3$ at nearly the same T=T-$T_C$. In the former, the soft mode is overdamped at this T, but the sharp central peak is clearly present and distinguishable from the broader phonon mode. These spectra suggest that there are two energy scales in the problem, the soft mode energy and the linewidth of the central peak. There have been many attempts to measure the linewidth by traditional neutron scattering[63], by neutron spin echo measurements[64], and by inelastic gamma-ray spectroscopy[65]. Thus far no one has succeeded in measuring a finite linewidth of the central peak and an upper limit of the energy linewidth of the central peak is 0.08 eV, or a relaxation rate of 20 MHz[65].

Another system where the central peak was extensively studied is the structural transition occurring at $T_c$=45 K in $Nb_3Sn$ as discussed above[38]. In this material there is an instability in the acoustic mode rather than an optic mode. Fig. 17 shows the spectra measured at a small wave vector. Clearly there is a mode softening with temperature and an increase in its intensity, but the change in the central peak intensity is more dramatic and dominates the spectra near the transition temperature.

In order to describe the observed spectra, an additional time response has to be included in the normal damped harmonic oscillator response function used to describe phonons in a solid. The inverse susceptibility is written

$$\chi^{-1}(qj,\omega) = \omega_\infty(qj)^2 - \omega^2 - i\omega\Gamma(\omega) \tag{5}$$

with a frequency dependent damping constant:

$$\Gamma(\omega) = \Gamma_0 + \frac{\delta^2}{(\gamma - i\omega)} \tag{6}$$

$\omega(qj)^2$ is the phonon energy with damping $\Gamma_o$. In Eq. (2) the additional frequency dependent damping represents another decay channel with inverse relaxation time $\gamma$ and coupling strength $\delta$. If $\gamma$ is much smaller that $\Gamma_o$ and $\omega_\infty^2, \delta^2, \omega_0^2$ then this response can be separated into a response for the central peak and one for the phonon. The observed intensity in a neutron experiment is:

$$S(q,\omega) = \frac{1}{\pi}(n(\omega)+1)\operatorname{Im}\chi(qj,\omega) \tag{7}$$

where $(n(\omega)+1)$ is the Bose-Einstein thermal occupation factor. With the above approximations, the integrated intensities can be written as:

$$\left.\frac{1}{\omega_0^2}\right|_{total} = \left.\frac{\delta^2(T)}{\omega_0^2(qj,T)\omega_\infty^2(qj,T)}\right|_{central} + \left.\frac{1}{\omega_\infty^2(qj,T)}\right|_{side} \tag{8}$$

where

$$\omega_o^2(qj,T) = \omega_\infty^2(qj,T) - \delta^2(T) \tag{9}$$

This functional form of (5) and (6) adequately describes the spectra observed in Fig. 24 as indicated by the lines through the data points. The derived quantities are shown in Fig.26. The quantity $\omega_0(qj)^2=0$ at $T_c$, which implies that $\omega_\infty(qj)^2 = \delta^2$ at $T_c$ from Eq. 9.

The key question is the origin of the additional decay channel. Is it an intrinsic feature or is it extrinsic related to some defects in the material? In fluids, there exist a central component associated with the density fluctuations. In this case the extra-low-frequency decay channel is provided by the internal degrees of freedom of the fluid and Mountain[66] described this by introducing a frequency dependent viscosity similar in form to the last term in Eq. 5. One of us suggested an intrinsic anharmonic mechanism related to differences between the collision-free and collision-dominated responses. The soft mode can interact with other phonons near the same wave-vector as well as other thermally generated phonons, which relax with time $\gamma^{-1}$. In these cases the central peak is dynamic and should have a measurable linewidth. However, no energy linewidth has been measured.

Other mechanisms related to defects have been proposed. This is based upon the observations that the central peak is sample dependent[68]. In a ferroelectric material, the central peak can even be made to disappear by annealing the sample[69]. In the defect mechanism the central peak is associated with the local distortion of the atoms about the defect[38,70]. This displacement field gives rise to diffuse scattering of x-rays or neutrons and is usually referred to as Huang scattering. This type of elastic scattering varies as $1/\omega(qj)^4$, which is predicted by Eq. 8. In normal materials Huang scattering is observed near the Bragg peaks because the long wavelength acoustic modes have very low energies. However, when there is a soft mode, as in the present case, and the impurities are of the right symmetry to couple to the soft mode, there can be an enhanced scattering at $\omega$=0 at the q-vector of the soft mode. If the defects are mobile, there is inelasticity associated with the central peak related to the motion of the defects and this could be a time scale longer than can be detected in the experiments[71]. Further support for an extrinsic origin comes from one experiment[72] which studied $SrTiO_3$ with different amounts of defects and showed an enhanced central peak intensity with an increase in defect concentration. Nevertheless in this experiment there was a large increase in the number of defects and a relatively small change in the central peak intensity while the similarity in the central peak intensity in insulators, metals and at magnetic phase transitions is difficult to understand in terms of defects because it is unlikely that these very different systems would have similar numbers of unknown defects.

The debate about the origin of the central peak continues to this day. The central peak is ubiquitous and has been observed in nearly all systems exhibiting structural phase transitions and is an important observation in hi-Tc superconductors.

**5.Two-length scale problem**

Another surprising observation in $SrTiO_3$ occurred 15 years after the discovery of the central peak. In his experiments to study the q-dependence of the critical scattering observed in $SrTiO_3$, Andrews[73] discovered that there were two length scales diverging as $T_c$ approaches. This was a surprising result since it violated the basic tenet of critical phenomena in phase transitions that there is a single length scale that becomes infinite at the transition temperature. This observation was repeated at other laboratories and observed in several other materials including magnetic transitions[74]. Several experiments found that the critical exponents associated with each length scale differed from each other. This problem appealed to Gen's scientific curiosity in that it was an unsolved observation in a system that he was very familiar with. He attacked this with the same enthusiasm and originality he used throughout his career.

Fig. 27 compares q-scans measured by x-rays and measured by neutrons[75]. The x-ray spectrum is decomposed into two parts, the narrow part with a long correlation length and the broader part corresponding to the shorter correlation length. Since the x-rays integrate over energy, the scattering includes both the phonon contribution and the central peak components of the scattering described in the previous section. In a high-resolution neutron scan, the spectrometer could be set for elastic scattering and then discriminates against the phonon contributions. The physics of the problem and the x-ray results challenged Gen to try to achieve a similar q-resolution in a neutron experiment to that obtained in the x-ray measurements. He was successful in that the resolution is only about a factor of 2 larger than the x-ray studies. With this resolution he should have seen a spectrum similar to the x-rays shown in Fig.27. The peak in the neutron experiment arises from the central peak observed in energy scans and the width is determined by the soft mode dispersion curves as discussed above.

Fig. 28 shows the temperature dependence of inverse correlation lengths[76] of the phonon part $\kappa_\infty$ and the central peak part $\kappa_c$. The quantity $\kappa_0$ is the inverse correlation length derived from q-dependence of the total scattered intensity; the central peak and the phonon contribution. $\kappa_0$ is the same as the broad scattering observed in the x-ray experiment shown by the dashed line in the figure. The dotted line labeled $\kappa_{L2}$ is the width of the narrow component measured in the x-ray experiment and <u>not</u> observed in the neutron study. Thus it was concluded that since the larger length scale is not observed in a neutron experiment but seen with x-rays it must be related to the differences in penetration depths of the two techniques.

Meanwhile, a similar feature was observed in some magnetic systems, which provided important clues to its origin in SrTiO$_3$. The longer length scale was observed in x-ray studies of magnetic critical scattering in Ho[77], NpAs[78] and UO$_2$[79] where the expected short range correlations exist along with the narrower component. Gen and his colleagues studied the magnetic scattering in single crystals of terbium[80-82] and holmium[77] in a series of novel experiments. Tb exhibits a spiral magnetic order below the transition temperature $T_s\sim229.3$ K in which the moments align ferromagnetically in the basal plane, but are rotated with respect to neighboring spins along the c axis by a turn angle $\phi$ of ~20° per layer. The resulting diffraction pattern then consists of magnetic satellite peaks offset from the nuclear reflections by the amount (0,0,±δ), where $\delta \sim 0.127$ Å$^{-1}$. By doing measurements at Q=(0,0, δ) the scattering angle is small (2θ$_s$=2.7°) and the transverse resolution is extremely high[80] (HWHM=0.00003Å$^{-1}$), better than that at a synchrotron x-ray experiment (HWHM=0.0005Å$^{-1}$). In addition, with a narrow neutron beam (300 μm) the sample can be scanned through the beam so the spatial dependence of the scattering can be measured. The schematic of the arrangement is shown in Fig. 29. The crystal is translated along the [00l] direction and scans are performed when the narrow beam traverses the edge of the sample or the center as shown in Fig 30. Both the narrow and the broad peaks are present in the scans, but the narrow component is 70% less than when the crystal traverses the edge of the sample. The suspected reason the narrow component had not completely disappeared is due to the beam hitting two faces perpendicular to [00l] as it enters and leaves the crystal. The next neutron experiment[82] performed was to more carefully mask the scattered beam so the skin regions of the crystal never reach the detector. The set up for this configuration is shown in Fig. 31 and the scans shown in Fig. 32. This is the same configuration used years earlier by Gen[83] to establish that the central peak in SrTiO$_3$ discussed above, originated from the bulk and not the surface of the material. The results in Fig. 32 show convincingly that when the beam hits the edge of the sample the narrow peak is present but absent in the bulk. Note the change in q-scale between the full scan shown in the insert and the scan in the bottom panel. The conclusion from the study of Tb is that the sharp component of the critical scattering is due to the outer skin of the crystal with a thickness of the order of 100μm.

The method of using narrow beams and translating the crystal perpendicular to the beam direction would not work for a conventional experiment for SrTiO$_3$ since the scattering angle for the R-point zone boundary is too large. One can reduce the scattering angle by going to shorter wavelengths, but this is more easily done using high energy x-rays generated by a synchrotron. With the high energy x-rays, the beam penetrates the sample and one can study bulk phenomena as well as the surface area. The desire to understand the narrow peak in $SrTiO_3$ led Gen to his collaboration with the German colleagues at the Hasylab in Hamburg. They used the same nearly perfect crystal that was grown for neutron experiments performed in the early 70's and discussed above. With a 100keV x-rays the sample scattering angle for the ½(1,1,5) position is 2.38° and the penetration of the beam is sufficient to study the bulk of the crystal. By moving the crystal[84] relative to the beam Shirane was able to probe the corner and the center of the crystal with very high q-resolution, $\sim 6 \times 10^{-5}$ Å$^{-1}$ along a given direction. The results are shown in Figs. 33a and b where the bulk and the corner are probed, respectively. The inset shows the region of the crystal being probed in each case. Fig. 33a shows only single Lorentzian with HWHM= $4 \times 10^{-3}$ Å$^{-1}$, which is consistent with the broad peak observed in the neutron and x-ray experiments at the same T-$T_c$. When the corner is probed as in Fig. 33b, the two-component lineshape is observed; the broader one as in Fig. 33a, and a sharp resolution limited peak originating from the skin. The experiment performed with higher resolution revealed a finite width, which was consistent with the first x-ray measurements of this second length scale. This result confirms that in $SrTiO_3$, as in Tb, the second length scale originates from the outer skin of the sample and not the bulk. Several more experiments[85] using high energy x-rays were performed with micrometer sized beams showed that the lattice parameter and the mosaicity and Bragg peak intensities varied with the distance from the surface. In addition a number of samples with different growth techniques and sample treatments were studied and the critical properties measured as a function of depth[86]. Moreover, a transmission electron microscope study showed that the dislocation density was much higher at the outer skin (100 m) than in the bulk[87]. The conclusion from these experiments is that the long-range strain fields in the vicinity of the surface seem to be responsible for the second length scale.

From these experiments it is clear that the sharp component is not an intrinsic property of the $SrTiO_3$, but is extrinsically related to differences between the bulk and the outer skin. There have been several attempts to understand why this sharp peak arises but the theories are not consistent with all the experimental results. They are all based upon the idea that the near surface region can transform at a higher temperature than the bulk and as the temperature is lowered the phase transition can propagate into the bulk. Most theories fail in describing the well-measured critical properties of the two length scales or the lineshapes. One theory[88] claims that the longer length scale is a consequence of the presence of quenched disorder near the surface and relies on an earlier theory, which showed that the transition temperature can be higher than the bulk and the critical exponents of the near surface region are different than the bulk[89]. Another proposal by one of us[74] suggests that random fields are created by defects introduced by surface preparation in the near surface region. A third explanation of the origin describes the coupling of the strain on the surface to the order parameter of the bulk. These surface strains can be considered as 'free', as opposed to the internal 'clamped' strain. The coupling to the order parameter fluctuations is weaker for free strains and thus the transition temperature is higher than the bulk. There is no agreement about whether the origin of the phenomena is associated with defects because it occurs in very different systems which would be expected to have very different defect densities at the surface: metals, insulators, structural phase transitions and magnetic phase transitions. Consequently no complete picture is available to explain all the experimental facts and the problem continues to be of interest today.

**6. Conclusions**

In this chapter we have discussed many of Gen's contributions to the study of structural phase transitions. They are a topic that interested Gen throughout his entire 50-year scientific career. Most of his unique and elegant experiments were performed with neutron scattering techniques at Brookhaven National Laboratory. We have summarised the results of his experiments showing that the soft mode concept does explain, at least qualitatively, much of the phenomena occurring at structural phase transitions. In some ways it is possibly pity that this work was acknowledged to be a triumph and led the condensed matter community to largely consider the subject as one that was solved. In our article we have attempted to show that there is more to be understood in the experimental results for one-dimensional metals, for the central peak and the two-length scale phenomena. For none of these examples is there a comprehensive theory that does justice to Gen's detailed experimental results.

We have tried to portray, by choice of topics presented, Gen's approach to doing science. He chose problems either by discussing the physics with his collaborators or by intuition, and he liked problems that were challenging and had an experimental solution. He was most excited about developing novel experimental arrangements that enabled him to test theoretical predictions or to test the experimental results of others that his intuition told him might be in error. He worked tirelessly in trying to solve an experimental problem and his skills with a triple axis spectrometer were unsurpassed. He was not bound by conventional spectrometer configurations and constantly reconfigured the instrument to enhance its capabilities and increase the important signal-to-noise.

Gen rarely worked alone and all of the work presented in this chapter is a result of collaborations with many individuals. We have consistently not referred to the collaborators in the text but they can be found in the list of references. We were fortunate to be among them as were many young scientists who came to Brookhaven to learn how to do experiments and to learn how to choose experiments. Gen was a competitor who liked to win whether he was doing physics, or playing tennis or playing poker. He set very high standards and expected his collaborators to work equally hard. He was however careful to ensure that they benefited from their experience at Brookhaven and what they achieved at Brookhaven would advance their careers and their reputation. His excitement and enthusiasm about science and neutrons was contagious and infected all who worked with him. He was a giant in the field and we shall all miss him.

**Acknowledgements**

We are grateful to many who worked on structural phase transitions at Brookhaven and especially to John Axe.

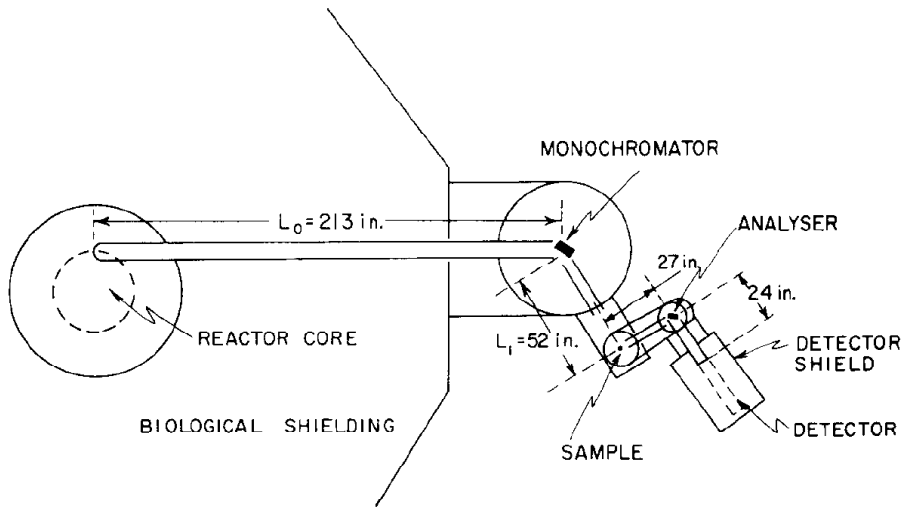

Figure 1 A schematic diagram of a triple axis spectrometer [From Ref. 8].

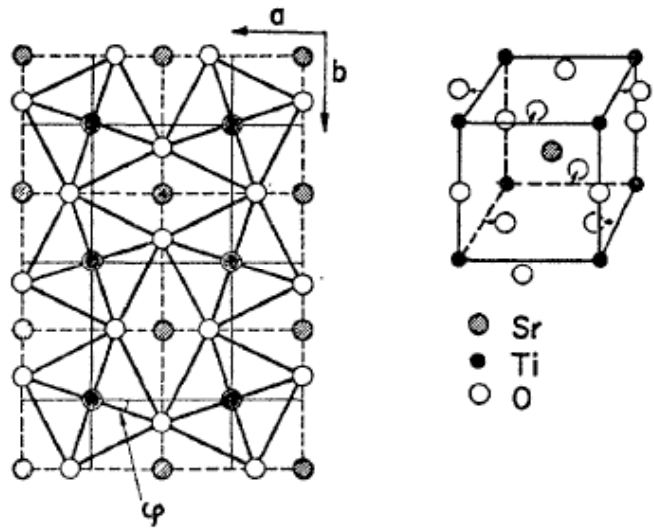

Figure 2 The structure of the distorted phase of $SrTiO_3$ proposed in ref. 11 [From Ref. 14]

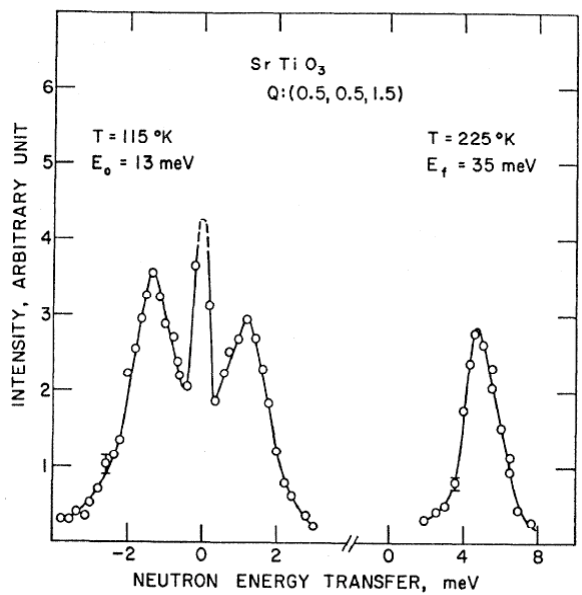

Figure 3 The neutron groups obtained for the (1/2,1/2,1/2) mode in SrTiO$_3$ at temperatures of 115K and 225K [From Ref. 14]

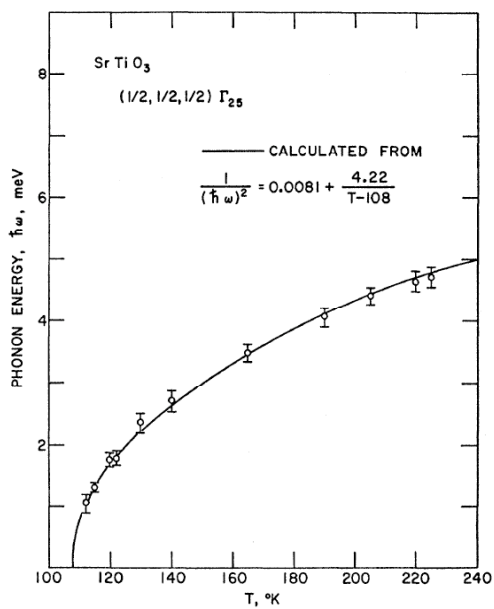

Figure 4 The temperature dependence of the triplet soft mode in SrTiO$_3$ showing good agreement with the predictions of Cochran [From Ref. 14]

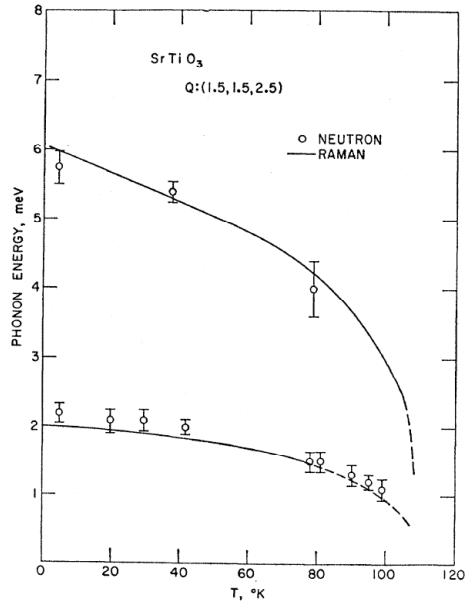

Figure 5 The temperature dependence of the frequencies of the singlet, $A_{1g}$ and doublet, $E_g$ modes below the transition temperature in $SrTiO_3$. There is good agreement with the frequencies observed by Raman scattering[13] [From Ref. 14]

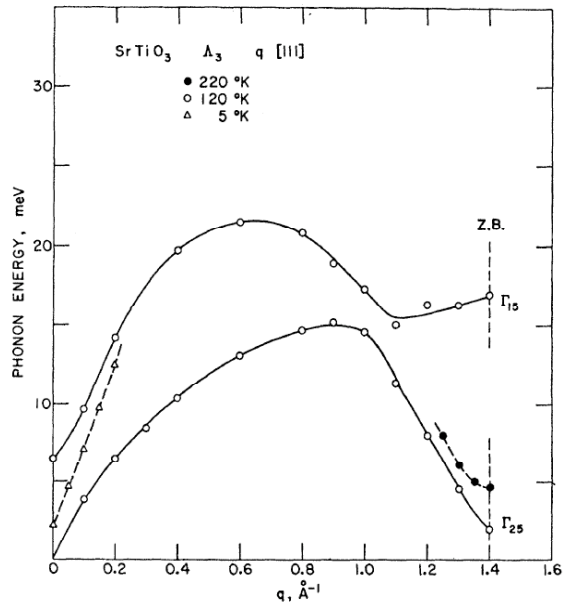

Figure 6 The dispersion curves in the [1,1,1] direction at different temperatures showing that the ferroelectric and antiferroelectric instabilities only modify the curves in a small region of reciprocal space [From Ref. 14]

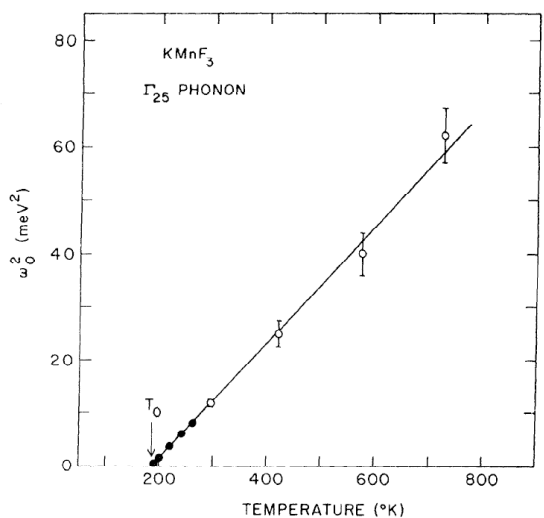

Figure 7 The temperature dependence of the soft mode in KMnF$_3$ [From Ref. 20]

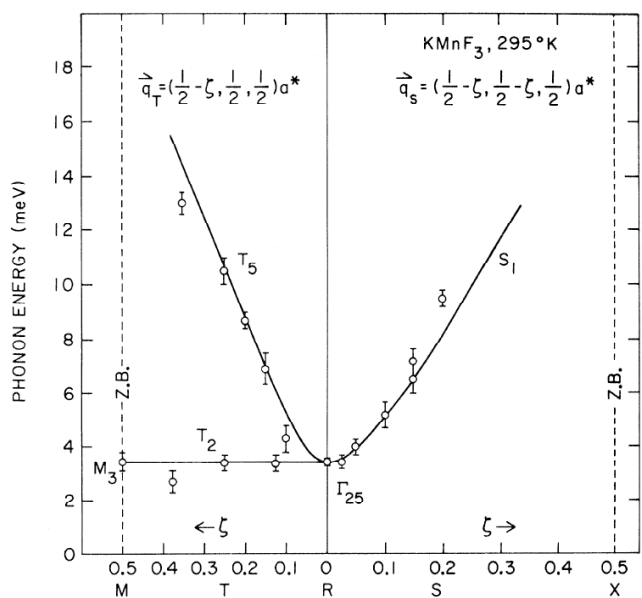

Figure 8 The dispersion relations for KMnF$_3$ at 295K showing the very flat and low frequency branch between the M and R points in the Brillouin zone [From Ref. 20]

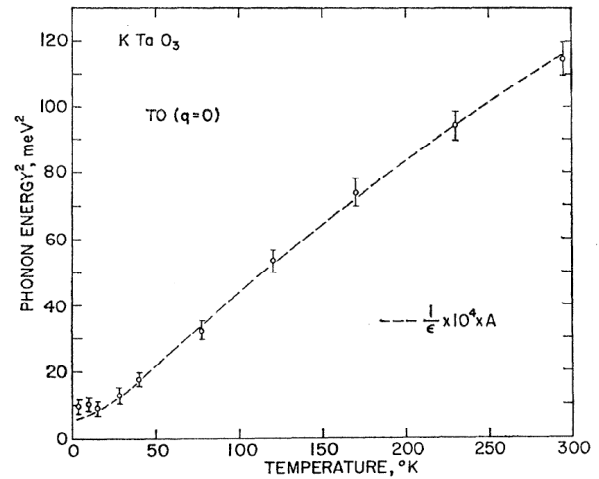

Figure 9 The temperature dependence of the square of the incipient ferroelectric soft mode frequency in KTaO$_3$ showing that it is inversely proportional to the dielectric constant [From Ref. 23]

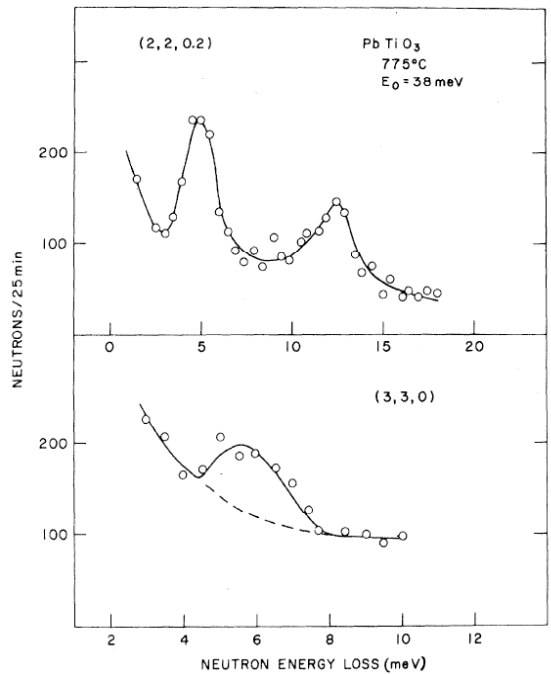

Figure 10 The neutron groups obtained for PbTiO$_3$ at a temperature of 1048K. At finite wave-vector there are well defined acoustic and optic peaks but at the zone centre the scattering from the optic mode is much less convincing. [From ref. 28]

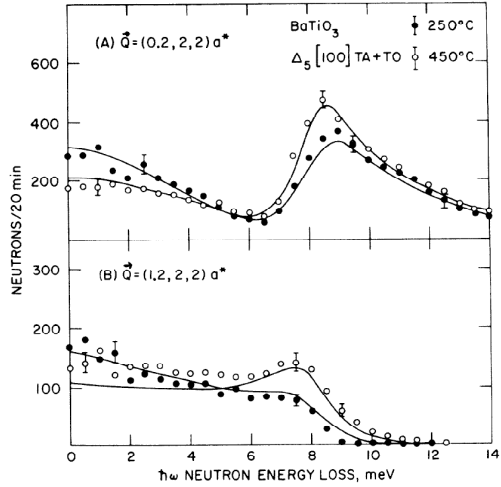

Figure 11 The interference effects between the TO and TA modes in BaTiO$_3$ resulting in different profiles in different Brillouin zones [From Ref 30]

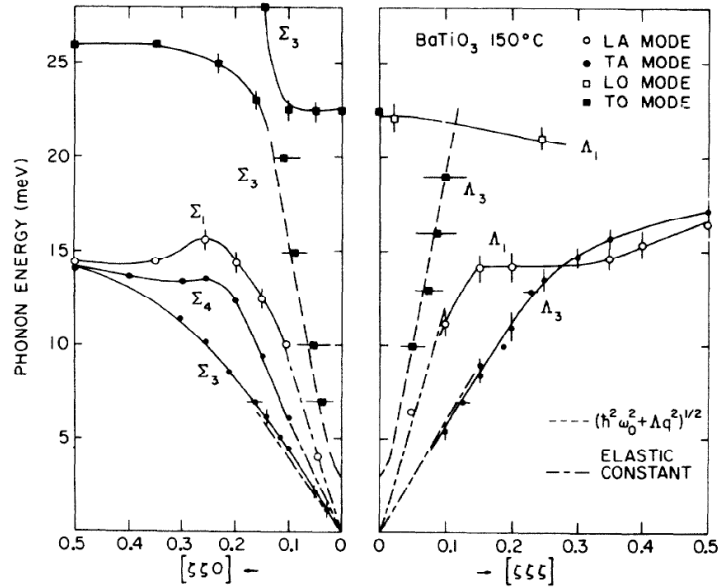

Figure 12 The low frequency dispersion relations in BaTiO$_3$ showing the absence of points at q=0 for the TO mode and the effects of the strong interaction between the TO and TA modes [From Ref. 30]

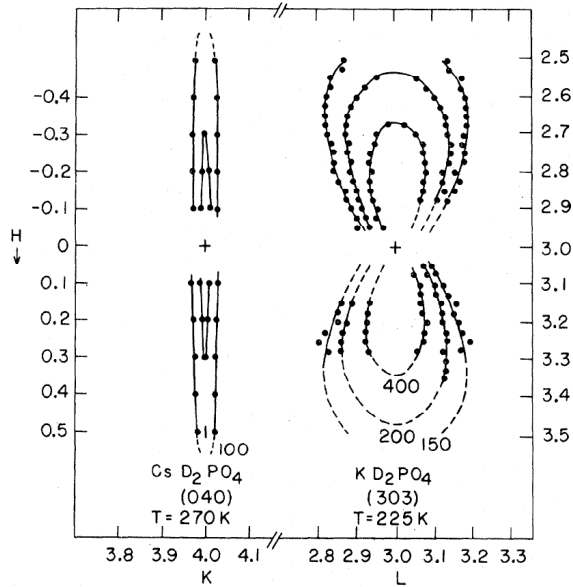

Figure 13. A contour map in the paraelectric phase of the quasi-elastic critical scattering from the (hk0) plane of $CsD_2PO_4$, where the ferroelectric axis is along the [010] and from the (h0l) plane of $KD_2PO_4$ where the ferroelectric axis is along the [001] [From Ref. 34].

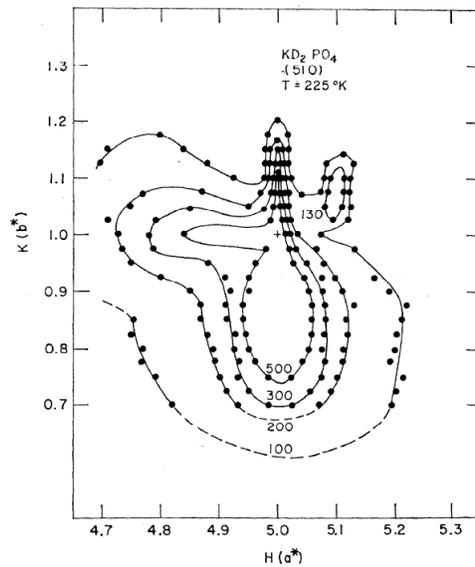

Figure 14. A contour map of the critical scattering in the (hk0) plane of $KD_2PO_4$ showing the effect of the linear coupling between the ferroelectric fluctuations and the acoustic modes [From Ref. 32].

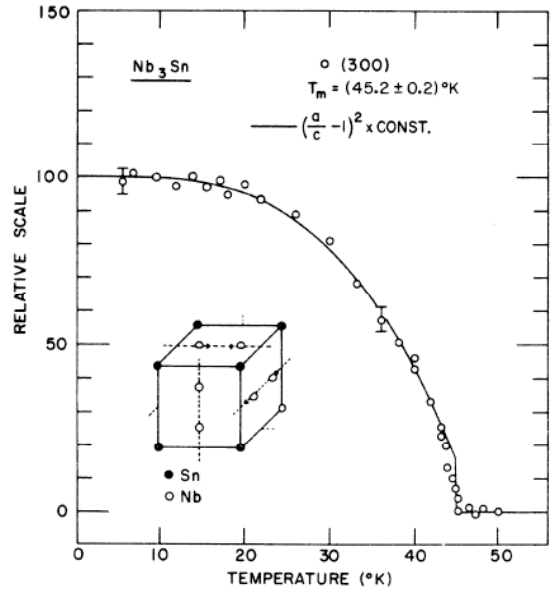

Figure 15. The temperature dependence of the (300) reflection in $Nb_3Sn$ that at high temperatures has a face centred structure [From Ref. 38]

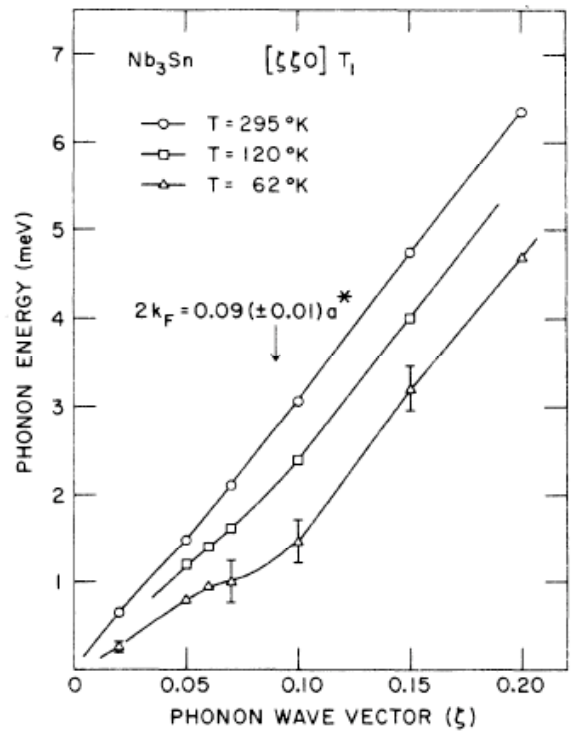

Figure 16. The dispersion relation for the $[\zeta\zeta 0]T_1$ acoustic mode as a function of temperature [From Ref. 38]

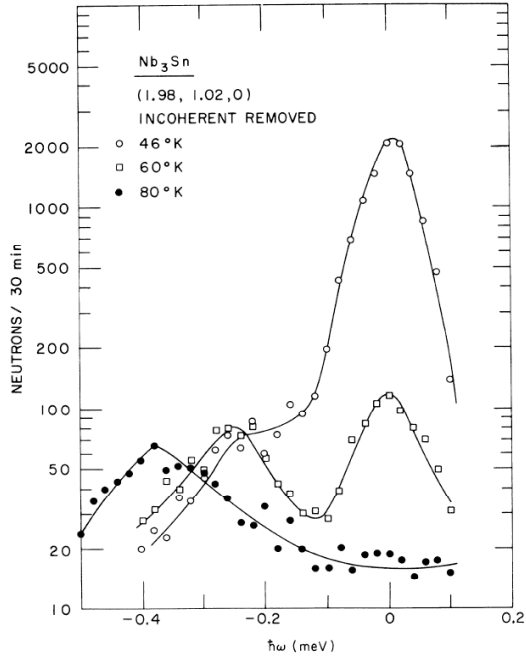

Figure 17. The scattering profiles observed for the $[\zeta\zeta 0]T_1$ acoustic mode as a function of temperature. The inelastic component softens more slowly than expected while the quasi-elastic component increases [From Ref. 38]

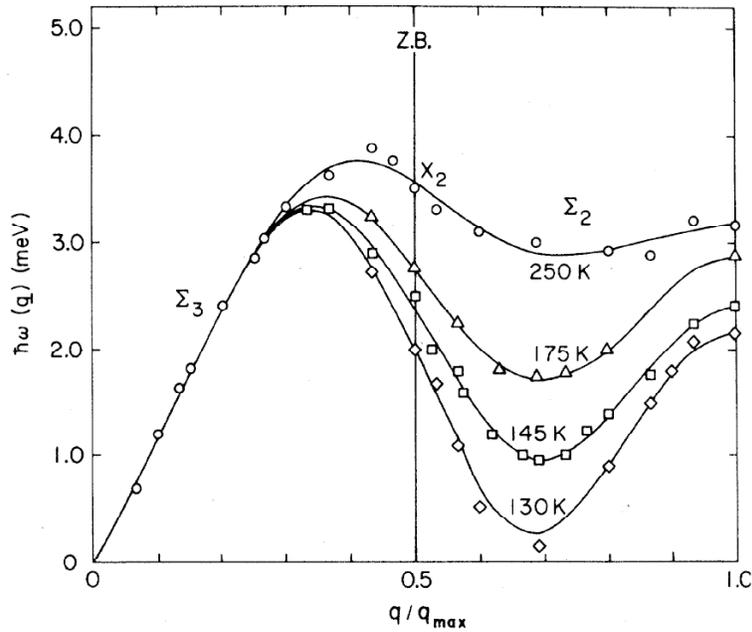

Figure 18. The dispersion relation of the soft mode in $K_2SeO_4$ along the [100] direction. The minimum in the frequency decreases as the phase transition is approached [From Ref. 42]

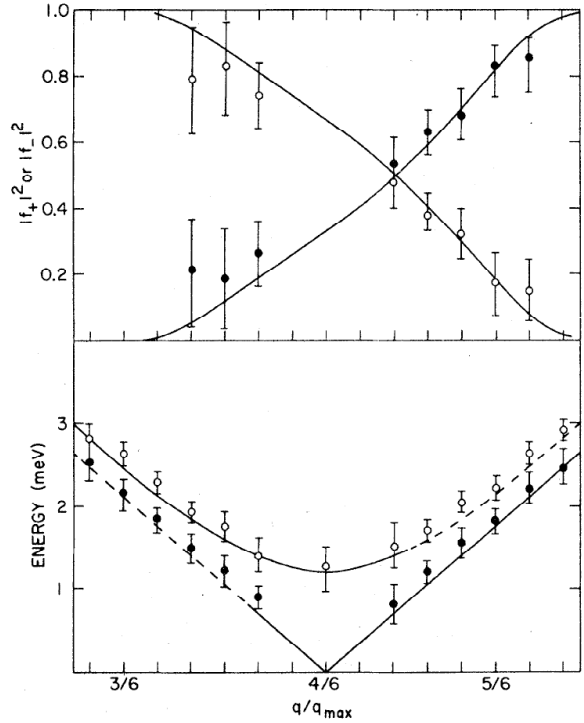

Figure 19. The dispersion relation at 40K in $K_2SeO_4$ showing a the frequencies and b the intensities for the scattering [From Ref. 42].

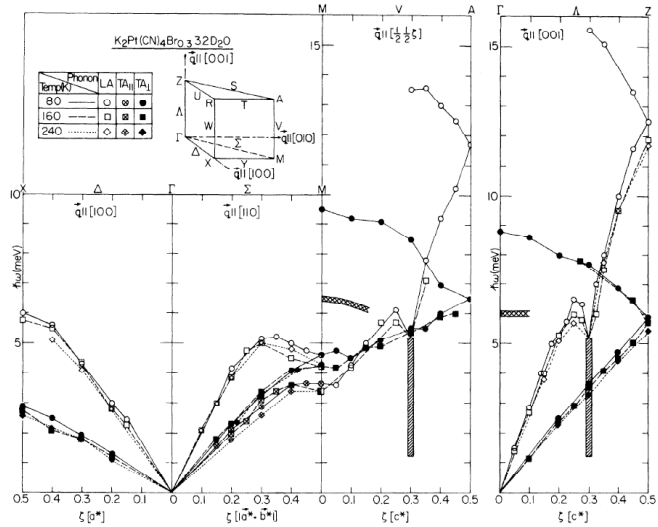

Figure 20. The phonon dispersion relations in KCP. Note the anomaly when the branches are propagating parallel to the [001] direction [From Ref. 45]

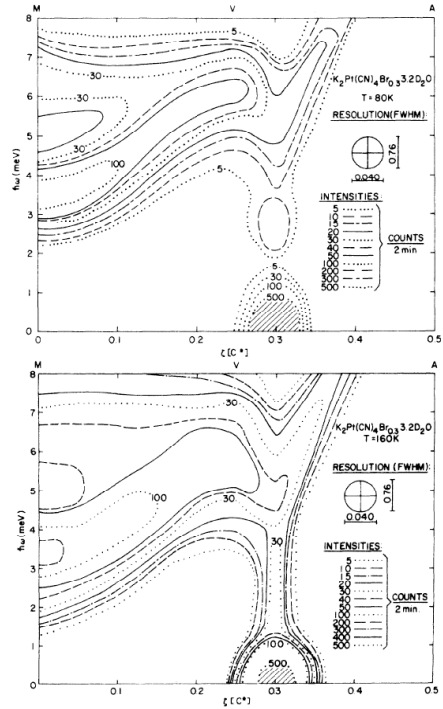

Figure 21. The intensity contours lose at the Kohn anomaly in KCP at 80K and at 160K giving detail about the intensity in the vertical part of the dispersion relation [From Ref. 45]

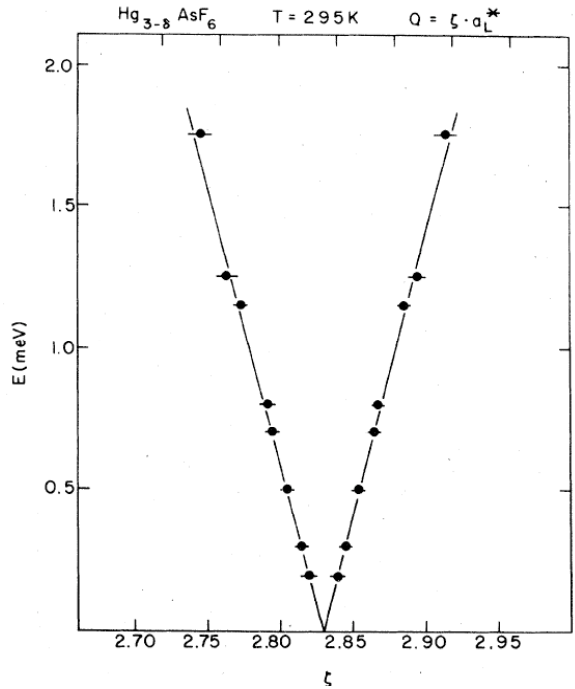

Figure 22. The peak values of the cross section obtained from constant energy scans across the Hg atom spectra in $Hg_{3-\delta}AsF_6$. The straight lines are the velocity of sound for these atoms [From Ref. 59].

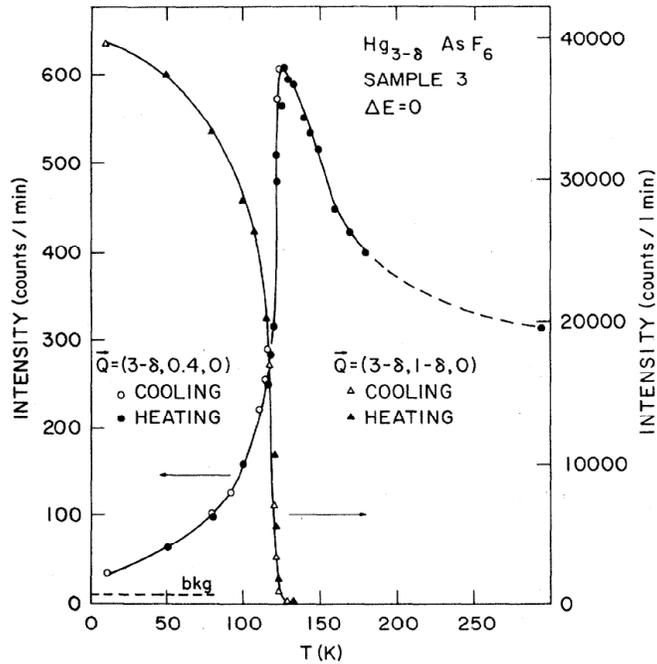

Figure 23. Temperature dependence of the elastic scattering at $(3-\delta, 0.4, 0)$ and at $(3-\delta, 1-\delta, 0)$ which are not and at the positions of low temperature Bragg peaks in $Hg_{3-\delta}AsF_6$ [From Ref. 57].

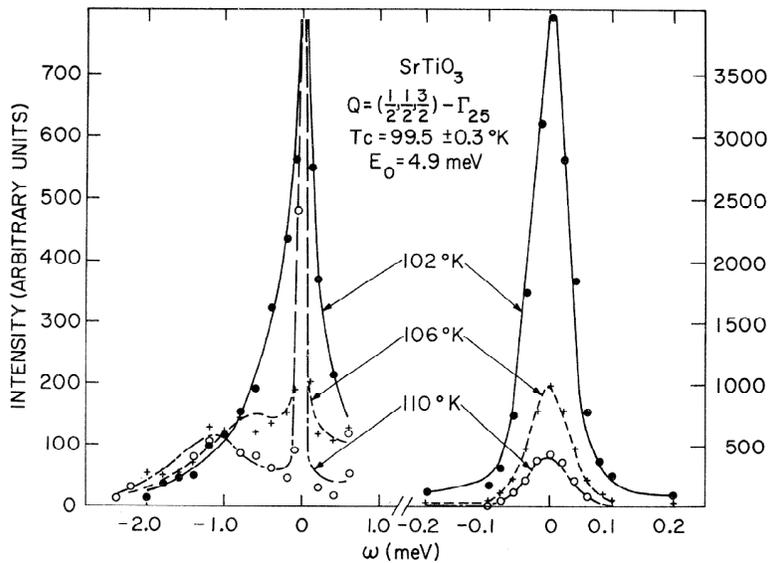

Figure 24. Inelastic spectra of $SrTiO_3$ measured at several temperatures. Left side shows the soft mode behavior and right side shows divergence of central peak [From Ref. 62].

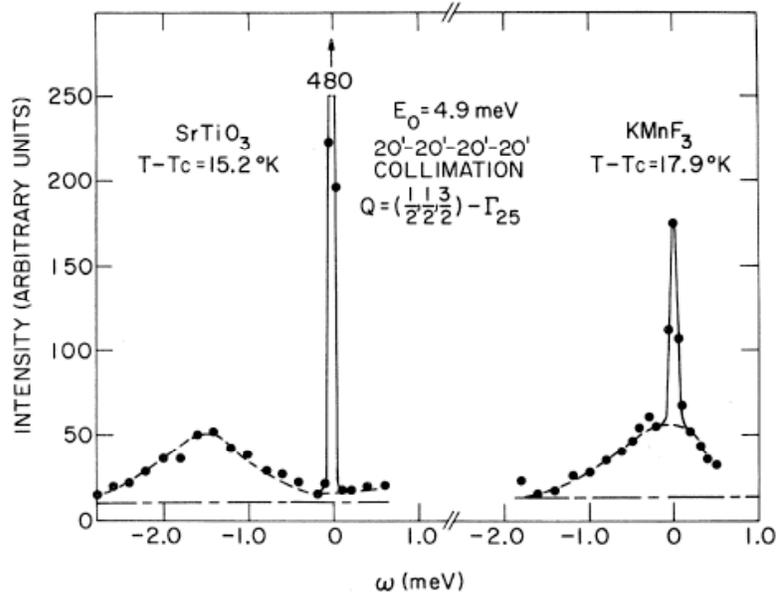

Figure 25. Inelastic spectra of SrTiO$_3$ (left) and KMnF$_3$ (right) at similar $\Delta T=T-T_c$. The incoherent scattering at $\omega=0$ has been subtracted [From Ref. 62]

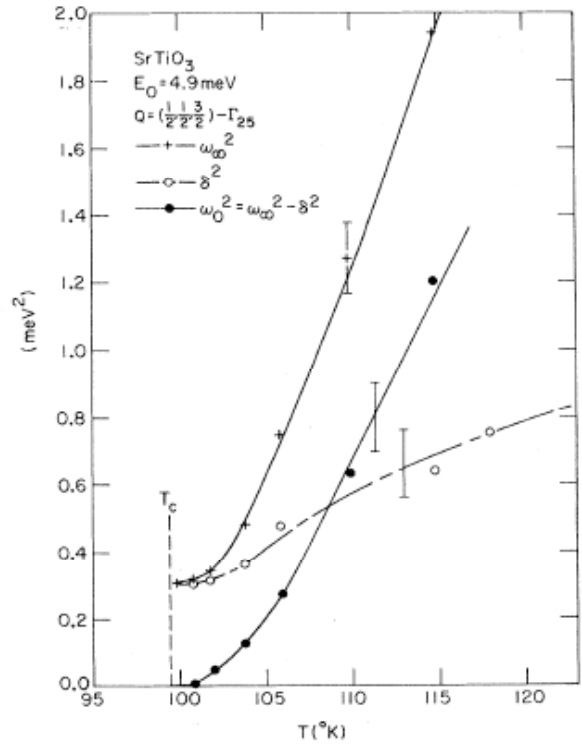

Figure 26. The quantities $\omega_\infty^2, \delta^2, \omega_0^2$ derived from fits of the data in Fig. 24 to Eqs 5-9 [From Ref. 62]

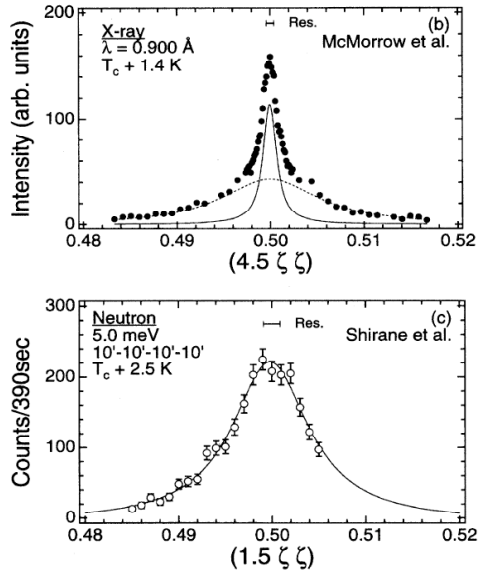

Figure 27. Top: X-ray scan [From D. F. McMorrow et al., Sol. St. Commun. 76, 443 (1990)] showing two length scales. The solid and dotted curves are fits to the data with Lonentzian squared and Lorentzian, respectively. Bottom: High resolution neutron diffraction scan. The solids line is a fit to a Lorentzian. [From Ref. 75]

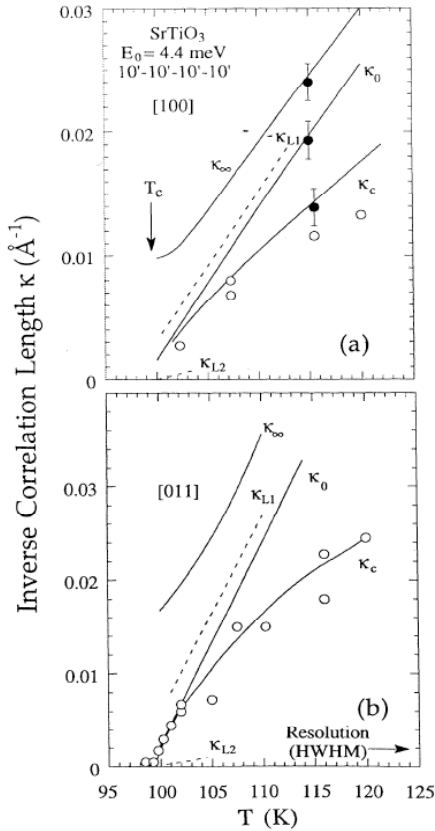

Figure 28. Temperature dependence of inverse correlation lengths measured by neutron scattering (circles and solid lines). Dashed lines are the two inverse correlation lengths measured by x-rays. [From Ref. 76]

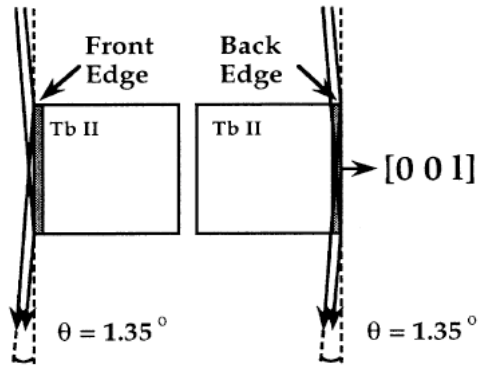

Figure 29. Scattering geometry for measuring the depth dependence of the scattering from Tb. [From Ref. 82]

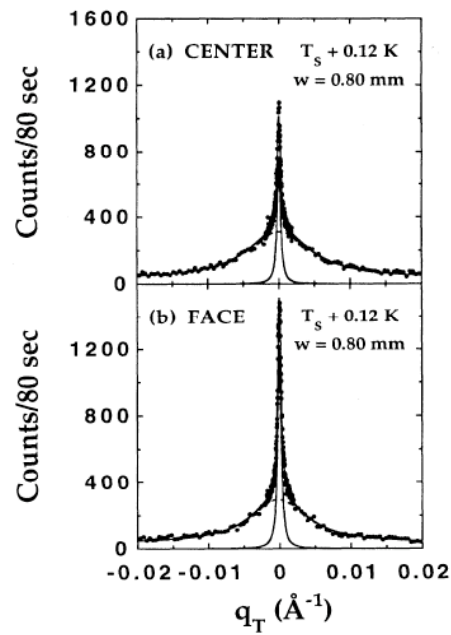

Figure 30. Transverse q-scans measured in Tb just above the transition temperature, $T_s$ for the beam traversing (a) the crystal center and (b) the crystal face. [From Ref. 80]

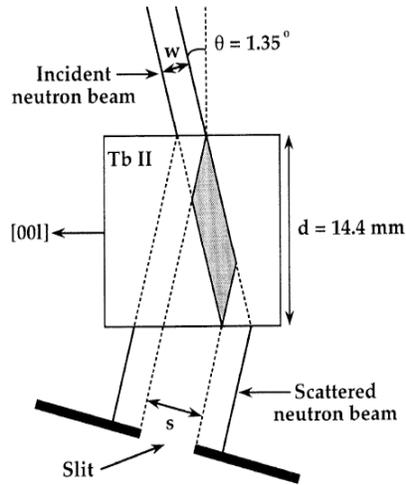

Figure 31. Schematic showing how scattered neutron beam is masked to eliminate contributions from entry and exit faces of Tb crystal. Only the shaded region contributes to the scattering [From Ref. 82]

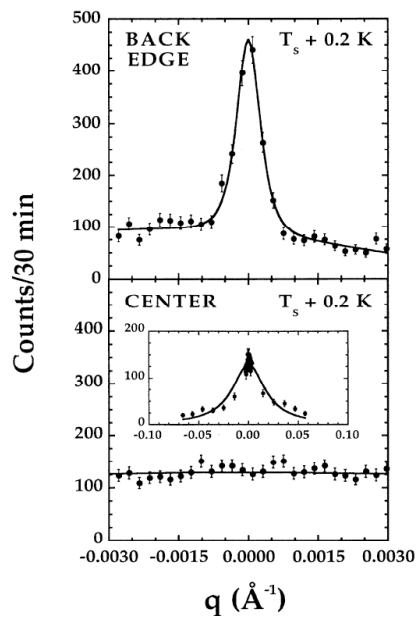

Figure 32. High resolution x-ray scans in Tb with the beam traversing (top) crystal face and (bottom) crystal center. Inset shows full range of scan. [From Ref. 82]

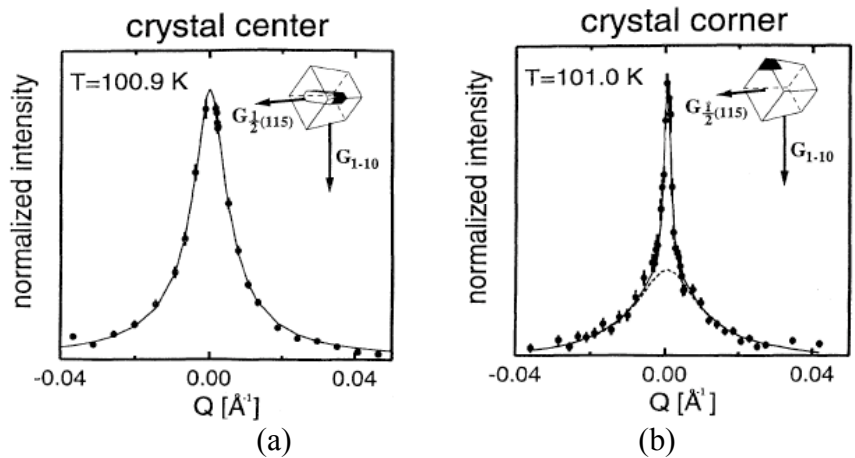

Figure 33. X-ray scans of $1/2(1,1,5)$ R-point along [011] direction with 100keV photon energy. Insets show (a) beam hitting the crystal center and (b) beam hitting the crystal corner. [From Ref. 84]